# Temperature Dependent Thermal Boundary Conductance of Monolayer $MoS_2$ by Raman Thermometry

Eilam Yalon[1], Özgür Burak Aslan[2], Kirby K. H. Smithe[1], Connor J. McClellan[1], Saurabh V. Suryavanshi[1], Feng Xiong[1,3], Aditya Sood[4,5], Christopher M. Neumann[1], Xiaoqing Xu[6], Kenneth E. Goodson[4], Tony F. Heinz[2], and Eric Pop[1,5,7,*]

[1]*Department of Electrical Engineering, Stanford University, Stanford, CA 94305, USA.* [2]*Department of Applied Physics, Stanford University, Stanford, CA 94305, USA & SLAC National Accelerator Laboratory, Menlo Park, CA 94025, USA.*[3]*Present address: Department of Electrical & Computer Engineering, University of Pittsburgh, Pittsburgh, PA 15261, USA.* [4]*Department of Mechanical Engineering, Stanford University, Stanford, CA 94305, USA.* [5]*Department of Materials Science & Engineering, Stanford University, Stanford, CA 94305, USA.* [6]*Stanford Nanofabrication Facility, Stanford University, Stanford, CA 94305, USA.* [7]*Precourt Institute for Energy, Stanford University, Stanford, CA 94305, USA.* [*]*E-mail: epop@stanford.edu*

**The electrical and thermal behavior of nanoscale devices based on two-dimensional (2D) materials is often limited by their contacts and interfaces. Here we report the temperature-dependent thermal boundary conductance (TBC) of monolayer $MoS_2$ with AlN and $SiO_2$, using Raman thermometry with laser-induced heating. The temperature-dependent optical absorption of the 2D material is crucial in such experiments, which we characterize here for the first time above room temperature. We obtain TBC ~ 15 $MWm^{-2}K^{-1}$ near room temperature, increasing as ~ $T^{0.65}$ in the range 300 – 600 K. The similar TBC of $MoS_2$ with the two substrates indicates that $MoS_2$ is the "softer" material with weaker phonon irradiance, and the relatively low TBC signifies that such interfaces present a key bottleneck in energy dissipation from 2D devices. Our approach is needed to correctly perform Raman thermometry of 2D materials, and our findings are key for understanding energy coupling at the nanoscale.**

## Introduction

Thermal interfaces are expected to dominate energy dissipation in 2D semiconductor devices and their characterization and understanding have become essential.[1-2] For example, drive currents in state-of-the-art 2D devices are critically limited by their heat dissipation capabilities[3-4] which are determined by thermal interfaces.[1] Moreover, understanding the fundamentals of heat flow across interfaces, namely the thermal boundary conductance (TBC), is an ongoing challenge in thermal physics of materials and calls for advances in existing experimental techniques.[5-6]

Among existing techniques, Raman thermometry is attractive to study 2D material thermal interfaces due to its material selectivity. Raman thermometry enables unprecedented (nm-scale) resolution along the laser path by simultaneously measuring the temperature of several Raman-active materials, even monolayers like graphene and h-BN.[7] Yet, characterization of thermal properties requires not only a measurement of the temperature but also an accurate definition of the input power density and a suitable thermal model. In Raman thermometry the temperature is measured optically, but the input power could be either electrical[1, 7-9] or optical.[10-14] The Joule input power in electrical heating experiments is well defined, but it requires fabricating high quality devices that carry high current densities sufficient to induce measurable Joule heating.[1] These requirements limit the materials and structures that can be used. The optical heating experiment, which simply requires increasing the Raman laser power applied to the sample, can readily be carried out on different materials and stacks. Nevertheless, a challenge of the Raman optical heating experiment is to accurately define the relevant input power and its density. Moreover, it is crucial to understand the heat dissipation mechanism in such experiments, as discussed below.

We previously measured the TBC of monolayer (1L) $MoS_2$ transistors on $SiO_2$ by Raman thermometry with direct electrical self-heating (during transistor operation), finding $G = 14 \pm 4$ $MWm^{-2}K^{-1}$ near room temperature in exfoliated and chemical vapor deposited (CVD) films.[1] Recently, Yasaei *et al.*[15] measured a larger value $G = 26 \pm 7$ $MWm^{-2}K^{-1}$, however by *indirect* heating across the $MoS_2$ from a Ti/Au heater on top. These values are within reasonable agreement, given the uncertainties of the measurements. Our previously measured TBC is equivalent to a Kapitza length $L_K \sim 90$ nm of $SiO_2$ (where $k_{SiO2} \approx 1.4$ $Wm^{-1}K^{-1}$) at room temperature, dominating the thermal resistance of $MoS_2$ devices.

Here we measure the TBC of monolayer CVD-grown $MoS_2$ on both $SiO_2$ and AlN via direct optical heating, as a function of temperature. Both interfaces ($MoS_2$-$SiO_2$ and $MoS_2$-AlN) show very similar values of TBC (~ 15 $MWm^{-2}K^{-1}$) in agreement with our electrical heating experiments.[1] $MoS_2$-AlN-Si test structures highlight the unique material selectivity of the Raman technique, allowing us to simultaneously measure the temperature of all three materials in the stack. In addition, the $MoS_2$-$SiO_2$ TBC is found to increase as $T^n$ ($n$ ~ 0.65) in the range 25 to 300 °C. This finding is not unexpected due to the (positive) temperature-dependence of the phonon specific heat, but is in contrast with a previous study which neglected the temperature dependence of the absorption.[11] Here we take into account the $T$-dependent absorption of the $MoS_2$, reporting it for the first time above room temperature. We also present a thermal model of the laser heating experiment for supported $MoS_2$, emphasizing that the measurement is sensitive to the TBC, but not to the thermal conductivity of the 2D film ($k_{2D}$) when the lateral thermal healing length is small compared to the laser spot size. Our findings provide important insights to understand heat transfer across 2D material interfaces, with implications for all optical, electronic, and thermoelectric devices based on such nanomaterials.

## Results and Discussion

Raman and temperature calibration

Figure 1 shows the measured $MoS_2$ films, their Raman spectra, and temperature calibrations. $MoS_2$ films were directly grown by chemical vapor deposition[16] (CVD, see Methods) onto thin $SiO_2$ (Figure 1a) and AlN[17] (Figure 1b) films on Si substrates. (To the best of our knowledge, this also represents the first demonstration of $MoS_2$ grown by CVD directly onto AlN.) Supporting Information Section 1 includes TBC measurements of exfoliated 1L $MoS_2$ showing similar results. The optical images in Figure 1a,b show large 1L triangular crystals (~ 50 µm on $SiO_2$ and ~ 30 µm on AlN) with small (~ 0.5 µm) bilayer regions.[18] All Raman measurements were carried out on the 1L $MoS_2$ areas, as verified by their Raman spectra (see Ref. 1).

The Raman spectra of $MoS_2$ on $SiO_2$ *vs*. stage temperature are shown in Figure 1c and the spectrum of $MoS_2$ on AlN at room $T$ is shown in Figure 1d. The temperature calibration of peak shift *vs*. stage temperature for $MoS_2$ $A_1$', Si LO, and AlN $E_2^2$ modes, which served as thermometers, are shown in Figures 1e,f,g. For our AlN film, the weaker $E_2^2$ mode is chosen as thermometer due to larger measured temperature coefficient ($\chi$ = 0.022 cm$^{-1}$/°C, where the mode frequency dependence is $\omega(T) = \omega_0 + \chi T$) compared with the $A_{1(TO)}$ mode ($\chi$ < 0.01 cm$^{-1}$/°C). The laser intensity was kept low ($P_{abs}$ < 20 μW) during the calibration to avoid measurable heating by the laser.

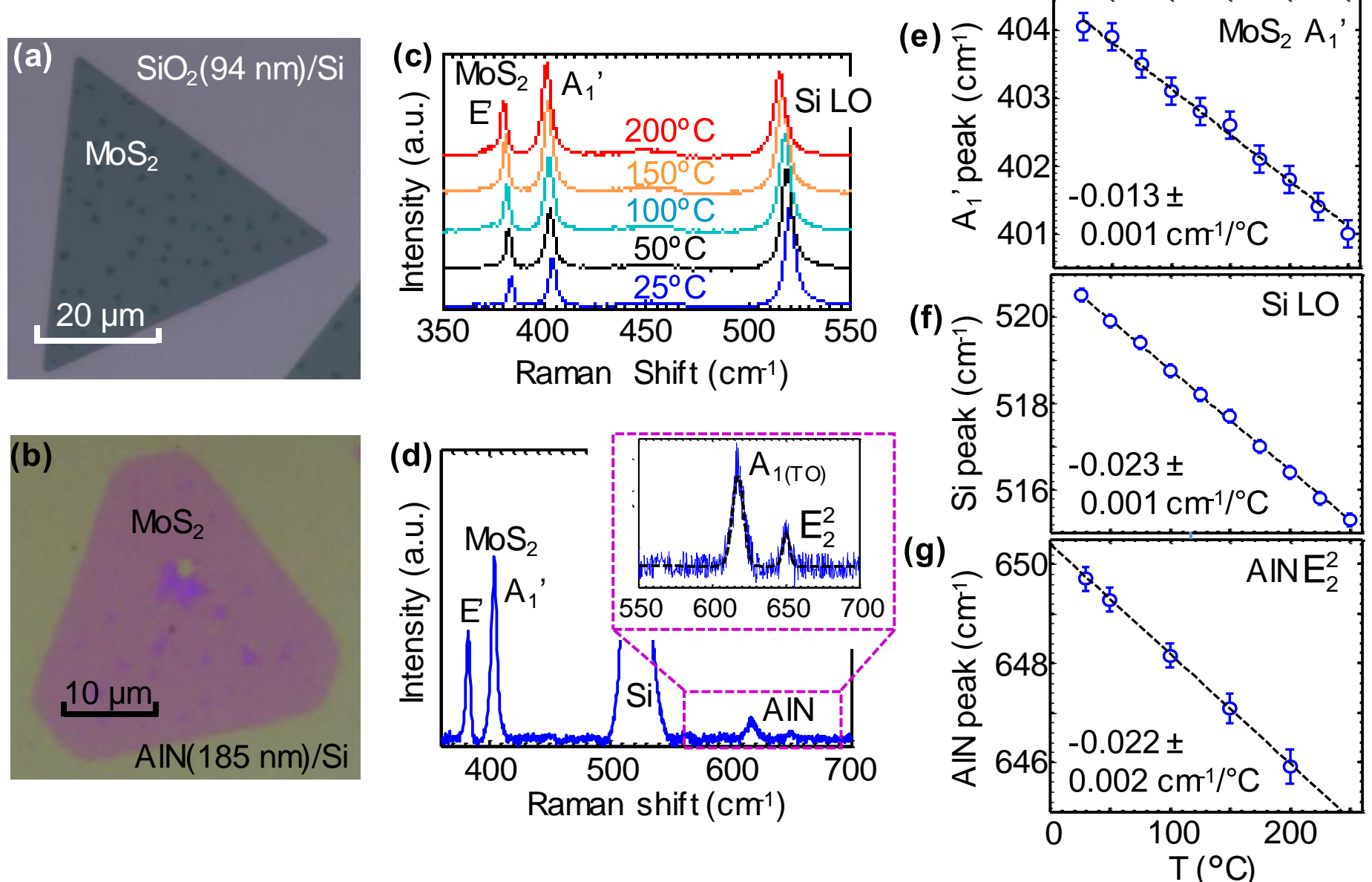

**Figure 1. Monolayer $MoS_2$ Raman and temperature calibration.** Optical image of CVD-grown $MoS_2$ on **(a)** $SiO_2$(94 nm)-Si and **(b)** on AlN(185 nm)-Si showing large triangular crystals and small bilayer spots (~0.5 μm in size). The measurements were carried out only on the 1L $MoS_2$. **(c)** $T$-dependent Raman spectra at varying stage temperatures of the 1L $MoS_2$ on $SiO_2$(94 nm)-Si. **(d)** Simultaneous Raman spectra of $MoS_2$ on AlN(185 nm) on Si(substrate). Inset shows the AlN peaks, which have a weaker Raman signal compared with the Si and $MoS_2$. Raman shift *vs.* stage temperature calibration of **(e)** $MoS_2$ $A_1$' peak, **(f)** the Si substrate peak, and **(g)** AlN $E_2^2$ peak. The absorbed laser power was kept below 20 μW in the $MoS_2$ to avoid measurable laser heating during the calibration.

<u>Laser heating</u>

To better understand the heat dissipation in our experiment we model the laser heating (schematic shown in Figure 2a) using a finite element thermal simulation (using COMSOL Multiphysics ®). The Fourier heat diffusion equation is solved with cylindrical coordinates (2D axisymmetric configuration) and a Gaussian shaped beam heat source. The simulated temperature rise in the sample is shown in Figure 2b. The bottom of the substrate is held at ambient temperature (isothermal boundary condition) and other surface boundaries are thermally insulating (adiabatic boundary condition). The thermal properties of the substrate are well known, including their temperature dependence around room temperature. The thermal conductivity of doped silicon can be expressed as $k_{Si} \approx 3\times10^4/T$ $Wm^{-1}K^{-1}$, the thermal conductivity of $SiO_2$ is $k_{SiO2} \approx \ln(T^{0.52}) - 1.6$ $Wm^{-1}K^{-1}$ (where $T$ is in Kelvin) and the TBC between the two is $G_{Si\text{-}SiO2} \approx 600$ $MWm^{-2}K^{-1}$ (following Refs. 1, 19-23). The heating profile is proportional to the laser spot size and the heat is dissipated mostly in the cross-plane direction into the Si substrate, typical for supported films as discussed below.

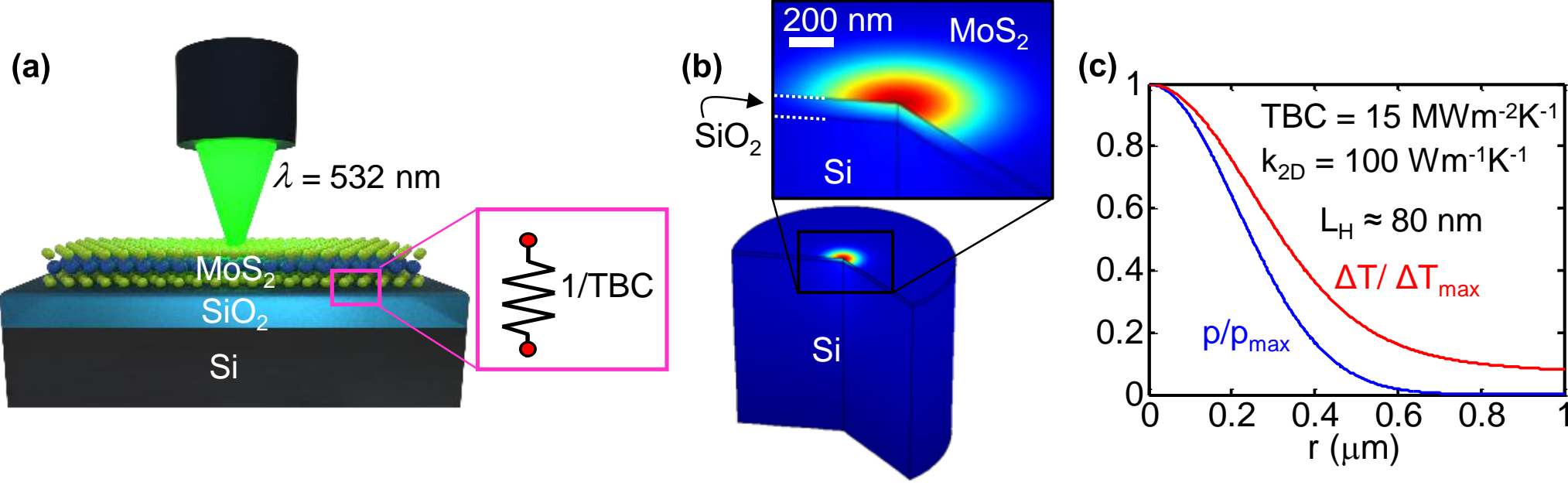

**Figure 2. Laser heating experiment and thermal model. (a)** Schematic sample structure and measurement setup. The thermal boundary resistance (= 1/TBC) at the $MoS_2$-$SiO_2$ interface is highlighted. **(b)** Simulated temperature rise during laser heating in 2D axisymmetric finite element model. **(c)** Normalized power density and temperature rise in the $MoS_2$ film *vs*. radial coordinate for $G = 15$ $MWm^{-2}K^{-1}$ and thermal conductivity $k_{2D} = 100$ $Wm^{-1}K^{-1}$ illustrating a small thermal healing length ($L_H \sim 80$ nm) compared to the laser beam size ($r_0 \sim 300$ nm). The short $L_H$ results in insensitivity of the measurement to the thermal conductivity of the 2D material (see Supporting Information Section 2).

The Raman laser heating technique was originally developed for bulk materials, where the thermal conductivity could be extracted,[24] and was later extended to suspended 2D films.[10, 25] In *suspended* structures the heat flows radially "in-plane" from the laser spot towards the supported part of the film, where the heat is sunk "cross-plane" (into the substrate). This combination of in-plane and cross-plane heat flow enables extracting both the thermal conductivity of the 2D film ($k_{2D}$) and the TBC with the supporting substrate. The separation between the two unknown parameters is obtained by varying the laser spot size.[10] By contrast, in *supported* 2D films typically the lateral thermal healing length is small compared with the laser spot size (see below), thus the heat flows predominately in the cross-plane direction into the substrate, and is therefore primarily sensitive to the TBC of the 2D film with the substrate.

The heat dissipation mechanism can be understood by comparing the heated area (laser spot) with the lateral thermal healing length $L_H = \sqrt{k_{2D}t_{2D}/g}$ (see e.g. Ref. 26) where $k_{2D}$ and $t_{2D}$ are the thermal conductivity and thickness of the 2D film, respectively, and $g$ is the *total* thermal conductance to the substrate per unit length. $L_H$ quantifies the characteristic length scale over which the heat travels laterally before sinking to the substrate and is on the order of ~ 80 nm for 1L $MoS_2$ on $SiO_2$(94 nm)-Si (assuming $k_{2D}$ ~ 100 $Wm^{-1}K^{-1}$,[11-12, 27] $t_{2D}$ = 0.615 nm, $g$ = 5 $MWm^{-2}K^{-1}$); much shorter than the laser spot ($r_0$ > ~300 nm). This is illustrated in Figure 2c which compares the simulated input power and the temperature profile in the $MoS_2$. Supporting Information Section 2 shows a similar comparison for different values of TBC and $k_{2D}$. We therefore set $k_{2D}$ = 50 $Wm^{-1}K^{-1}$ in our thermal model,[12, 27] and validated that the extracted TBC values were insensitive (within measurement uncertainty) to changes of $k_{2D}$ in the range 20 to 120 $Wm^{-1}K^{-1}$. Given these inputs to the finite element thermal model, the $MoS_2$-$SiO_2$ TBC remains a single fitting parameter

We note that the laser spot size dominates the uncertainty of the measurement and it is useful to vary the spot size during the heating experiment by sweeping the z-position of the objective in order to reduce that uncertainty (~10% in spot radius, equivalent to ~20% in spot area, see Supporting Information 3). The uncertainty in spot size is comparable to the $L_H$, further signifying the insensitivity of the measurement to the in-plane thermal conductivity.

Absorbed laser power

The temperature in our experiment is measured by converting the Raman peak shifts during laser heating to temperature, using the calibrations shown in Figure 1. In addition to the temperature measurement, the extraction of TBC requires the characterization of absorbed laser power in the supported $MoS_2$ film, including its temperature dependence. The absorbed power is obtained here by multiplying the incident laser power by the absorption of a free-standing 1L $MoS_2$ and by the enhancement factor of the substrate, as discussed below. We note that the optical absorption of $MoS_2$ as a function of temperature is measured and reported here for the first time over the 25 to 300 °C temperature range.

Due to the strong interferences of multiple light reflections in the 1L $MoS_2$-$SiO_2$-Si structure, it is difficult to directly measure the absorption of the supported film. We therefore estimate the absorption of our 1L $MoS_2$ on $SiO_2$-Si by multiplying the free-standing 1L $MoS_2$ absorption by a wavelength-dependent factor: The intensity of the electric field (of the electromagnetic wave of the laser) at the top surface of $SiO_2$-Si substrate relative to the intensity of the *incident* electric field (namely, the enhancement factor of the $SiO_2$-Si stack). The free-standing absorption is obtained by measuring the (*T*-dependent) absorption of 1L $MoS_2$ on quartz, where the quartz dielectric function is known (see Methods). The enhancement factor is calculated by the transfer matrix method[28] using the refractive indices of the materials in the stack, which are known at room temperature, and we assume that these reflections do not change significantly with *T*.

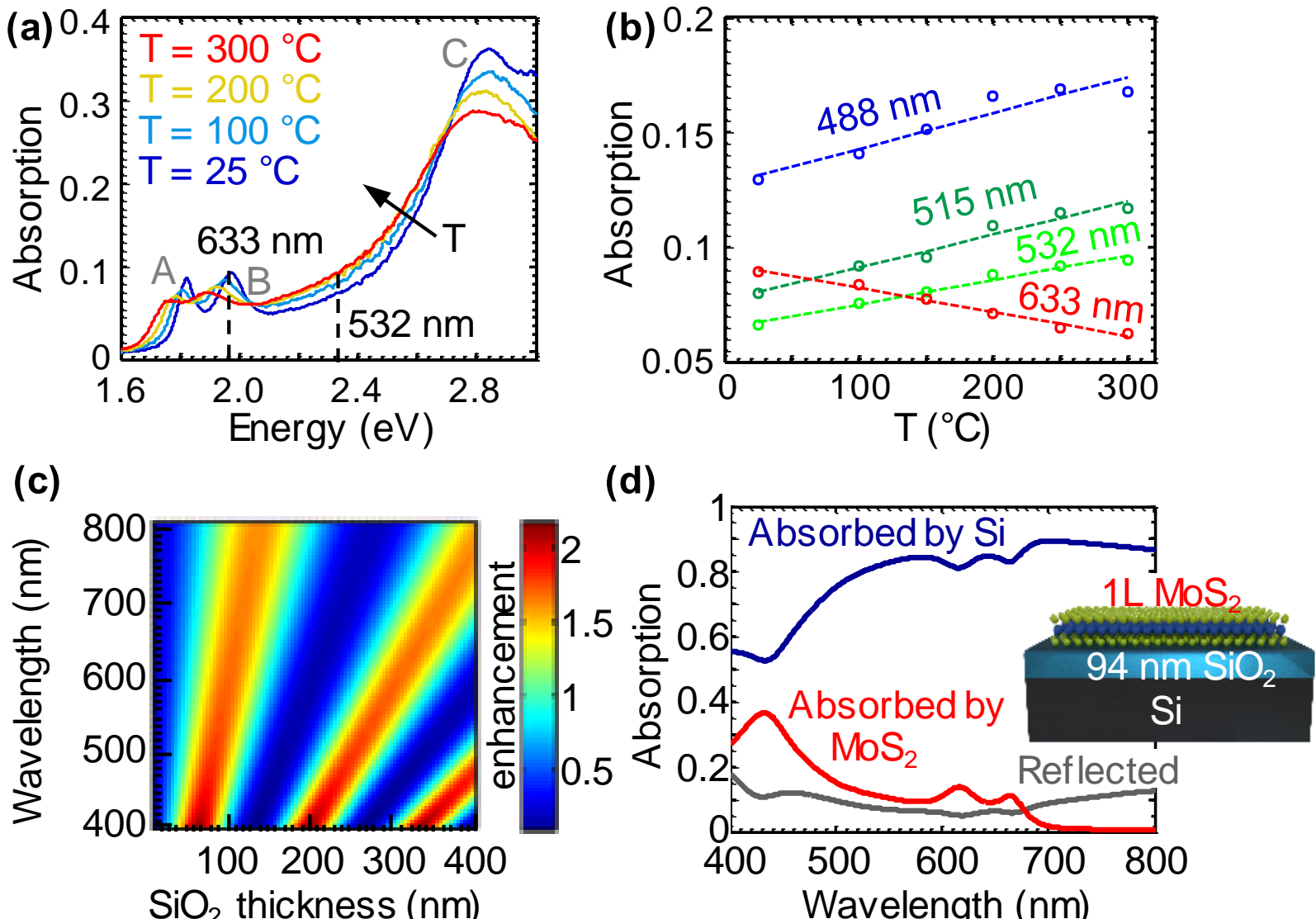

**Figure 3. Optical absorption of monolayer $MoS_2$. (a)** Absorption of free-standing 1L $MoS_2$ *vs*. laser energy at temperatures from 25 to 300 °C. Two typical visible Raman laser lines are indicated by dashed black lines (532 and 633 nm), and the excitonic peaks are labeled. **(b)** The temperature dependence of optical absorption in free-standing 1L $MoS_2$ (markers) and linear fits (dashed lines) for typical Raman laser lines: 488, 515, 532, and 633 nm. Note that the absorption can change by more than ~ 30% between room temperature and ~ 250 °C. **(c)** Colormap of the substrate enhancement factor, defined as the intensity of the electric field (on $SiO_2$-Si substrate) relative to the intensity of the incident electric field for laser wavelengths in the range of 400 to 800 nm and varying $SiO_2$ thickness in the range of 10 to 400 nm. **(d)** Calculated absorption spectra of 1L $MoS_2$ on $SiO_2$(94 nm)-Si (red), the Si substrate (dark blue) and the fraction of the light reflected (gray). Dielectric functions used for the calculation in (d) are taken from Ref. 29. The absorbed laser power in a supported 2D film is calculated as $P_{abs} = P_{in}\alpha_f E$, where $P_{in}$ is the incident laser power, $\alpha_f$ is the absorption of the free-standing film and $E$ is the enhancement factor of the substrate.

Figure 3a displays four selected absorption spectra at selected stage temperatures between 25 and 300 °C. Both A and B excitons redshift from ~ 1.82 and 1.97 eV to ~ 1.75 and 1.88 eV, respectively. Such redshifts with increasing temperature are typically found due to reduced overlap in the orbitals forming the band in the thermally expanded crystals. Figure 3a also shows that the absorption changes significantly near the A and B excitons. The 1L $MoS_2$ absorption is obtained by averaging over two measured samples at commonly used laser wavelengths of 488, 515, 532, 633 nm and fitting them to a linear function of $T$ (dashed lines in Figure 3b). It is evident that in green lasers, for instance, the absorption at 250°C is increased by ~ 30% compared with its room temperature value. Overlooking this ~30% increase in the absorbed power can result in underestimation of the TBC (and thermal conductivity in suspended films) of the 2D material in Raman thermometry experiments. These temperature-dependent absorption results are essential for calculating the absorbed laser power of 1L $MoS_2$ for Raman thermometry or other optoelectronic applications above room temperature.

Figure 3c shows the calculated enhancement factor of the $SiO_2$ (with thickness $t_{ox}$) on Si substrate and Figure 3d shows the absorption spectra of 1L $MoS_2$ and the Si substrate, as well as the reflected light of the $MoS_2$-$SiO_2$(94 nm)-Si stack. Most of the laser power is absorbed in the Si substrate (within an absorption depth, here ~ 0.65 µm at wavelength $\lambda$ = 532 nm) heating it above the ambient temperature in spite of its relatively large thermal conductance (see Supporting Information Section 4).

Thermal boundary conductance and its temperature dependence

Figure 4 summarizes the results of the measured TBC of 1L $MoS_2$ with $SiO_2$ and AlN. The temperature rise *vs*. *absorbed* power in the $MoS_2$ on $SiO_2$ (thickness $t_{ox}$) on Si, with $t_{ox}$ = 31 nm (blue) and 94 nm (red) is shown in Figure 4a. The laser spot size ($r_0$) is characterized in Supporting Information Section 3 and the absorbed power is calculated as described in Figure 3. The error bars in measured temperatures are from the uncertainty in Raman measurement and peak fitting, and the error bars of absorbed power are obtained by propagating error from uncertainty in the three factors mentioned earlier: incident laser power, absorption, and enhancement factor.

The temperature measured by Raman thermometry is defined as follows:[10]

$$T_m = \frac{\int_0^\infty T(r) \cdot r \cdot e^{-r^2/r_0^2} dr}{\int_0^\infty r \cdot e^{-r^2/r_0^2} dr} \tag{1}$$

where $r_0$ is the beam radius. The TBC acts as a single fitting parameter in the thermal model discussed earlier and it is extracted for each measured temperature, given the input power and spot size (for details on the measured beam radius see Supporting Information Section 3).

The extracted TBCs are presented in Figure 4b for different $MoS_2$ temperatures. The stage temperature varied between 24 and 200 °C. The results in Figure 4b include samples with and without a thin $AlO_x$ capping layer (see Methods), showing no measurable difference. The error bars of the TBC are obtained from the uncertainty in the measured thermal resistance, which is a function of the measured temperature, calculated absorbed power, and the measured spot size. The $MoS_2$-$SiO_2$ interface accounts for more than 50% of the thermal resistance of the $MoS_2$-$SiO_2$(94 nm)-Si stack, and more than 70% for the stack with 31 nm $SiO_2$. The magnitude and uncertainties of the Si-$SiO_2$ interface thermal properties[19,21] are small compared with the error bars in our measurement. The $MoS_2$-$SiO_2$ TBC shows a weak increase with temperature over the measured range, from 20 to 300 °C. The black dashed line is given by $G = 0.37T^{0.65}$ (in $MWm^{-2}K^{-1}$, with $T$ in K), and the TBC increases by ~ 40% in the measured temperature range.

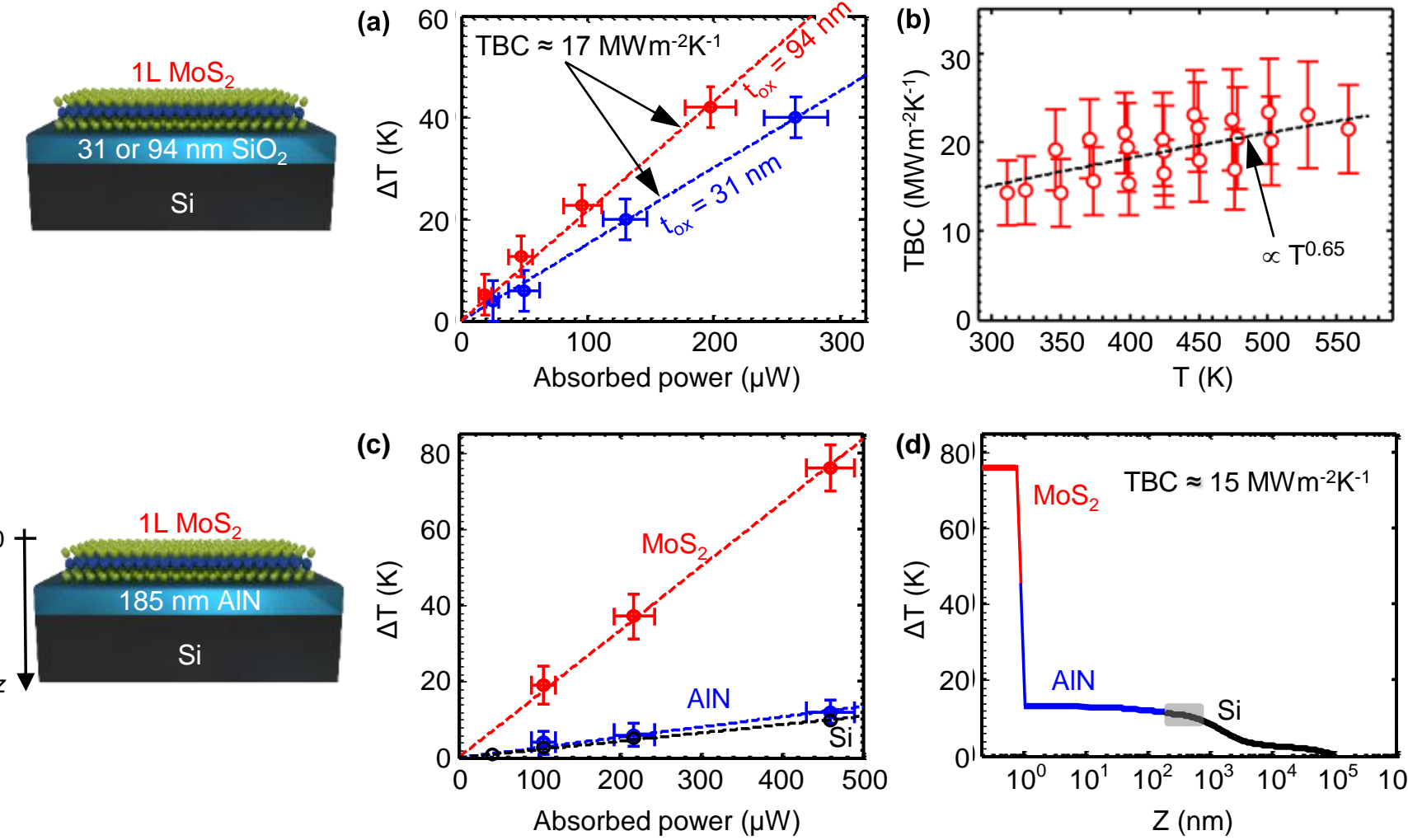

**Figure 4. TBC of monolayer $MoS_2$, and its temperature dependence. (a)** Measured temperature rise of 1L $MoS_2$ on 31 nm (blue) and 94 nm (red) $SiO_2$ on Si substrate *vs*. absorbed laser power. Calculated absorbed power takes into account the measured temperature-dependent absorption of $MoS_2$ (Figure 3a,b) and reflections from the substrate, given the exact oxide thicknesses (Figure 3c). Slope of dashed lines is the thermal resistance of each sample.

**(b)** Measured TBC *vs*. $MoS_2$ temperature, weakly increasing up to ~ 300 °C. The extracted $MoS_2$-$SiO_2$ TBC is 17 ± 5 $MWm^{-2}K^{-1}$ (averaged between 300 and 400 K) and the dashed black line shows $T^n$ dependence with $n = 0.65$. **(c)** Measured temperature rise of $MoS_2$ (red) on AlN (blue) and Si (black) as a function of absorbed power in the $MoS_2$. The temperature of each material in the stack is measured independently by its Raman signal. The temperature rise of the Si is due to the absorbed power in the Si (see Figure 3d and Supplemental Information Section 4). **(d)** Simulated temperature rise (Gaussian weighted across laser spot size) in the sample along the z-axis, at absorbed power of 480 μW. The gray area represents the absorption depth of the highly doped Si substrate, where the Si temperature is measured (and where power is absorbed). The extracted $MoS_2$-AlN TBC is 15 ± 4 $MWm^{-2}K^{-1}$ and the estimated thermal conductivity of the AlN film is $k_{AlN}$ = 60 ± 15 $Wm^{-1}K^{-1}$.

We emphasize here the importance of measuring the temperature-dependent absorption of the $MoS_2$. Supporting Information Section 5 shows that the thermal resistance with respect to the *incident* laser power increases with temperature which might lead one to conclude that the TBC decreases with increasing temperature,[11] a physically unlikely result, as we discuss below. The reason for this apparent increase of the thermal resistance is the increase in absorption of the optical power with temperature (for the 532 nm laser, Figure 3a-b). By taking this effect into account we find that the thermal resistance with respect to the *absorbed* power shows a minor decrease with temperature and hence the TBC slightly increases with temperature as shown in Figure 4b.

The TBC typically follows the temperature dependence of the specific heat, which at low temperatures ($T \ll \Theta_D$, where $\Theta_D$ is the lower Debye temperature of the two materials forming the interface) takes the form $\propto T^{d/m}$ where $d$ is the dimensionality of the material with phonon dispersion $\omega \sim q^m$, $q$ being the phonon wavevector (see Ref. 2). At high temperatures ($T > \Theta_D$) the specific heat approaches a constant, and the TBC likewise is expected to saturate. Thus, at intermediate temperatures, in the transition between these two regimes, the TBC can be expected to increase more weakly, as $\propto T^n$ (where $0 \leq n < d/m$).[5, 30] In the case of 2D materials, the graphene-$SiO_2$ TBC was studied in the range of 50 to 500 K and showed an increase with $T$ up to ~ 300 K, where the TBC saturates.[31-32] In light of the higher $\Theta_D$ of graphene[33-34] than $MoS_2$ we expect that the TBC of $MoS_2$-$SiO_2$ should have a weak temperature dependence for intermediate temperatures as well, which is what we observe experimentally (Figure 4b).

We also examined the TBC of monolayer $MoS_2$ with AlN, as shown in Figures 4c,d. The crystalline AlN layer has a measurable Raman signal, whose temperature dependence can be calibrated as shown in Figures 1d,g. The absorbed power in the $MoS_2$ is obtained as outlined earlier, with the relevant enhancement factor calculated for the AlN(185 nm)-Si stack (Supporting Information Section 6). The measured temperatures of the $MoS_2$ (red), AlN (blue) and Si (black) in the Raman laser heating experiment are shown in Figure 4c as a function of the absorbed power in the $MoS_2$ film.

The differential temperature measurement of all three materials in the stack allows us to estimate the $MoS_2$-AlN TBC as well as the thermal conductivity of the AlN film. In this case the $MoS_2$-AlN TBC is obtained from the temperature difference across the interface and $k_{AlN}$ is obtained from the AlN temperature (see note in next paragraph on the AlN-Si TBC). This emphasizes the exceptional capability of the Raman technique to measure thermal interfaces without prior knowledge of the thermal properties of the materials in the stack, thanks to its material selectivity. The dashed lines in Figure 4c represent the thermal resistance of $MoS_2$ and AlN in the laser heating experiment calculated from the model presented in Figure 2, with TBC = 15 $MWm^{-2}K^{-1}$ and $k_{AlN}$ = 60 $Wm^{-1}K^{-1}$, with a laser spot radius of $r_0$ = 200 nm (measured for the objective used in this experiment, magnification 100×, N.A.=0.9). The simulated temperature profile along the z-axis when the absorbed power in the $MoS_2$ is $P_{abs,MoS2}$ = 480 μW is shown in Figure 4d.

We note that the temperature rise in the Si is not due to the absorbed power in the $MoS_2$, but rather due to the power absorbed by Si itself ($P_{abs,Si}$ = 6.2 mW) within its absorption depth ~ 0.65 μm at the 532 nm laser wavelength[35] (gray area in Figure 4d). In the Si substrate, both the absorbed power (heating) and the Raman signal (measured $\Delta T$) originate from the Si surface. The absorbed power decays exponentially with the absorption depth, whereas the substrate is 500 μm thick. The uncertainty in the measured thermal conductivity of the AlN is relatively large, since its temperature rise is low relative to the uncertainty in Raman temperature measurement. Yet the measured thermal resistance is dominated by the $MoS_2$-AlN TBC and therefore the uncertainty in the measured TBC is comparable to the one measured in $MoS_2$-$SiO_2$ in spite of the relatively large error in the measured $k_{AlN}$. The simulated temperature profile shown in Figure 4d is insensitive (< 3% error) to the AlN-Si TBC (in the range TBC > 10 $MWm^{-2}K^{-1}$) since the AlN is mostly heated by the Si substrate in this case.

An important result presented in Figure 4 is that the measured TBCs between monolayer, CVD-grown $MoS_2$ and two different materials: 1) amorphous $SiO_2$, and 2) crystalline AlN are very similar. It is interesting to note that the TBC between 1L $MoS_2$ and hBN also showed similar values (~ 17 $MWm^{-2}K^{-1}$) in a recent study.[36] Furthermore, the measured TBC values of $MoS_2$-$SiO_2$ and $MoS_2$-sapphire interfaces reported in Ref. 15 are within a comparable range. These findings suggest that the TBC is dominated by the material with the weaker energy irradiance across the interface (here $MoS_2$) as was recently proposed in Ref. 5. The energy irradiance is proportional to the specific heat and (cross-plane) carrier velocity, and is expected to be lower in $MoS_2$ compared with hBN, AlN and $SiO_2$.[37-39]

The lower value of the TBC obtained here and in Ref. 1 (by *direct* optical or electrical heating of the $MoS_2$) compared with the TBC reported in Ref. 15 (the $MoS_2$ being heated *indirectly* by a metal on top) could be due an "internal thermal resistance"[40-42] between high-frequency optical phonon (OP) modes and low-frequency modes. If the TBC is dominated by low-frequency modes the interfacial thermal transport has two main contributions: (i) an internal thermal resistance between the OP modes that are excited (electrically or optically) and the low-frequency modes, and (ii) an external thermal resistance between the low-frequency modes in the $MoS_2$ and the substrate. An indirect heating experiment probes only the latter (external) contribution to the thermal resistance, while the result obtained here, in a direct heating experiment, is the relevant one for devices, where the power is dissipated within the $MoS_2$ film or device.

The Raman thermometry method with optical heating can also be applied, in principle, to multilayer (ML) films if the temperature-dependent absorption and Raman shifts of the specific ML can be determined. We can also estimate the internal TBC *between* individual layers of a bulk (or thick ML) film, by normalizing the cross-plane bulk thermal conductivity of $MoS_2$ (~2 W/m/K)[43] by the layer thickness (~0.615 nm). The internal TBC between $MoS_2$ layers is thus ~3 $GWm^{-2}K^{-1}$, two orders of magnitude higher than the TBC with the substrate (for comparison, the interlayer thermal conductance of bulk graphite[2] is ~18 $GWm^{-2}K^{-1}$). In other words the $MoS_2$-$SiO_2$ TBC is equivalent to ~200 layers, or a Kapitza length corresponding to ~130 nm thick bulk $MoS_2$. However, in thinner $MoS_2$ films quasi-ballistic cross-plane transport effects[37] must be considered to properly account for the interplay of the TBC and that of the internal thermal resistance, which could be the subject of future work.

## Conclusions

In summary, we measured the temperature-dependent TBC of 1L $MoS_2$-$SiO_2$, and the TBC of $MoS_2$-AlN by Raman thermometry with optical heating. We identified some critical points in the analysis of the laser heating experiment: 1) understanding the heat dissipation which is dominated by cross-plane transport across the interface; 2) characterization of the absorbed power density, including measurement of the $T$-dependent absorption and measurement of the laser spot size at varying offsets of the focal plane. Near room temperature, we obtain similar values of the 1L $MoS_2$-$SiO_2$ TBC as previously measured by electrical heating, equivalent to a Kapitza length of ~ 90 nm $SiO_2$. Knowledge of the $T$-dependent absorption $\alpha(T)$ is essential to extract the correct $T$-dependent thermal properties in such optical heating experiments. Taking into account the measured $\alpha(T)$ we find that the TBC weakly increases with temperature in the range 20 to 300 °C, in contrast to a previous study.[11]

We characterized the TBC of $MoS_2$-AlN by leveraging the simultaneous temperature measurement of all three materials in the stack ($MoS_2$-AlN-Si), as uniquely enabled by the Raman technique. The obtained $MoS_2$-AlN TBC is similar to that of the 1L $MoS_2$-$SiO_2$ measured here and 1L $MoS_2$-hBN measured in an earlier study,[36] suggesting that the TBC of these interfaces is limited by the $MoS_2$, which has lower (cross-plane) Debye temperature and phonon irradiance. Our findings are essential to interpret Raman thermometry experiments, and to understand the heat dissipation in all opto-electronic devices based on 2D materials.

## Methods

### Material Growth

We study monolayer (1L) $MoS_2$ grown by chemical vapor deposition (CVD) on $SiO_2$ ($t_{ox}$ = 31 or 94 nm), as well as AlN (185 nm), both on Si substrates ($p^+$, electrical resistivity of 1-5 mΩ·cm). A subset of the $MoS_2$-$SiO_2$(94 nm)-Si samples were capped by ~ 15 nm $AlO_x$, as described in Ref. 1. More details can be found in Ref. 16. The 185 nm thick AlN was grown by MOCVD on the Si substrate immediately after an HF dip to remove the native oxide. The dislocation density in the AlN film is expected to be ~ $10^9$ $cm^{-2}$ due to the lattice mismatch and polarity difference.[17]

The $MoS_2$ was deposited by CVD at 850 °C directly on the AlN, in a process similar to the one described in Refs. 16 and 18.

Characterization

Raman spectroscopy was carried out using a Horiba LabRam Revolution HR instrument with a 532 nm laser, 1800 l/mm grating and two different objectives: a 100× long working distance (LWD) objective with numerical aperture N.A. = 0.6, and 100× objective with N.A. = 0.9. Temperature calibration was done with a Linkam THMS600 stage in air ambient. Peak position of each Raman mode was fitted to a single Lorentzian line shape. After calibrating Raman peak shifts vs. temperature on a hot stage (Figure 1) we increase the applied laser power and the $MoS_2$ temperature was measured by converting the Raman peak shifts (of $MoS_2$, Si and AlN) to temperature rise (the measured temperature is a weighted Gaussian across the laser spot).[10]

The absorption of CVD-grown 1L $MoS_2$ on a quartz substrate was measured at temperatures from 25 °C up to 300 °C in the Linkam THMS600 stage. We studied two samples for all temperatures, and calculated the absorption spectra of free-standing 1L $MoS_2$ from these measurements.[44-46]

## Acknowledgements

We thank Pawel Keblinski for fruitful discussions. We acknowledge the Stanford Nanofabrication Facility (SNF) and Stanford Nano Shared Facilities (SNSF) for enabling device fabrication and measurements. This work was supported in part by National Science Foundation (NSF) EFRI 2-DARE grant 1542883, by the NSF Center for Power Optimization of Electro-Thermal Systems (POETS) under grant EEC-1449548, by the NSF DMREF grant 1534279, by the Stanford SystemX Alliance, by NSF DMR-1411107 (O.B.A.) and by the Department of Energy, Office of Science, Basic Energy Sciences, Materials Sciences and Engineering Division, under Contract DE-AC02-76SF00515 (T. F. H.). E.Y. acknowledges partial support from Ilan Ramon Fulbright Fellowship and from Andrew and Erna Finci Viterbi Foundation. K.K.H.S. acknowledges partial support from the Stanford Graduate Fellowship (SGF) program and NSF Graduate Research Fellowship under Grant No. DGE-114747.

## Additional information

Supporting information is available in the online version of the paper. TBC of exfoliated $MoS_2$; Sensitivity of laser heating to thermal conductivity of supported 2D film; Laser spot; Heating by absorbed power in Si substrate; Thermal resistance of incident vs. absorbed power; Enhancement factor of AlN-Si substrate.

Reprints and permissions information is available online at. Correspondence and requests for materials should be addressed to E.P.

# Supporting Information

## Temperature Dependent Thermal Boundary Conductance of Monolayer $MoS_2$ by Raman Thermometry

Eilam Yalon[1], Özgür Burak Aslan[2], Kirby K. H. Smithe[1], Connor J. McClellan[1], Saurabh V. Suryavanshi[1], Feng Xiong[1,3], Aditya Sood[4,5], Christopher M. Neumann[1], Xiaoqing Xu[6], Kenneth E. Goodson[4], Tony F. Heinz[2], and Eric Pop[1,5,7,*]

[1]*Department of Electrical Engineering, Stanford University, Stanford, CA 94305, USA.* [2]*Department of Applied Physics, Stanford University, Stanford, CA 94305, USA & SLAC National Accelerator Laboratory, Menlo Park, CA 94025, USA.*[3]*Present address: Department of Electrical & Computer Engineering, University of Pittsburgh, Pittsburgh, PA 15261, USA.* [4]*Department of Mechanical Engineering, Stanford University, Stanford, CA 94305, USA.* [5]*Department of Materials Science & Engineering, Stanford University, Stanford, CA 94305, USA.* [6]*Stanford Nanofabrication Facility, Stanford University, Stanford, CA 94305, USA.* [7]*Precourt Institute for Energy, Stanford University, Stanford, CA 94305, USA.* [*]*E-mail: epop@stanford.edu*

## 1. TBC of exfoliated $MoS_2$

Figure S1 compares the thermal boundary conductance (TBC) of monolayer (1L) $MoS_2$-$SiO_2$, prepared by chemical vapor deposition (CVD, black) and exfoliation (red). Evidently, there is no measurable difference between the CVD and exfoliated samples. We note that the CVD samples were directly grown onto the $SiO_2$-Si substrate (at 850 °C, see Ref. 1 ) while the exfoliated samples were prepared at room temperature from bulk $MoS_2$ on identical substrates. These results suggest that any residual strain from the high-temperature growth of $MoS_2$ has little, if any, effect on the TBC of this material with $SiO_2$.

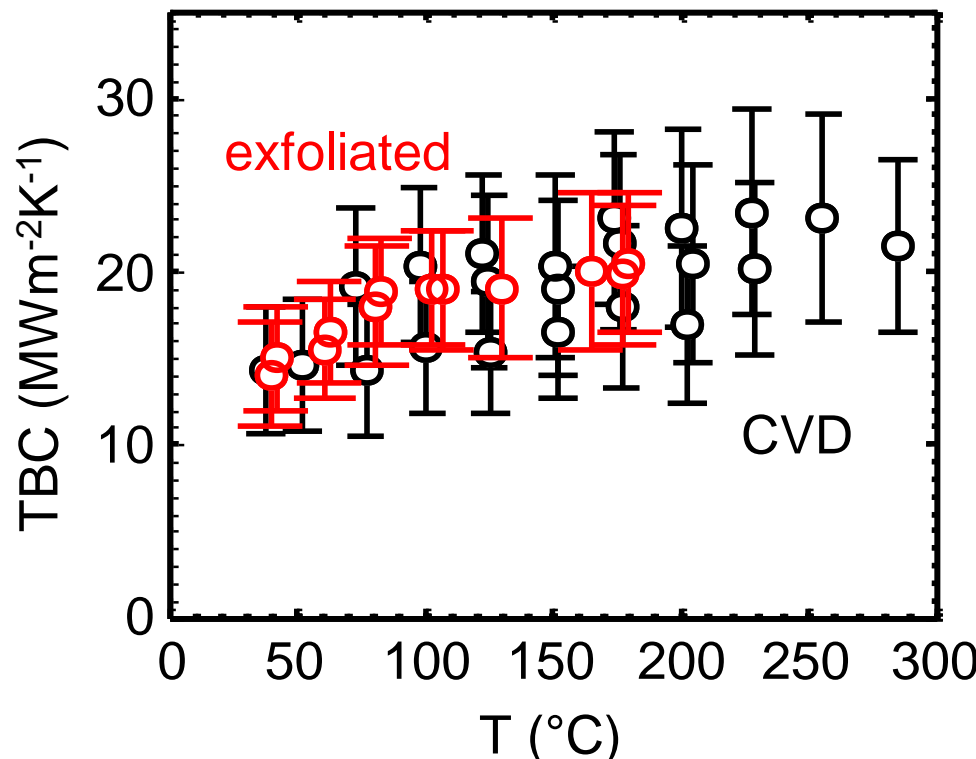

**Figure S1 | Temperature dependent thermal boundary conductance (TBC) of exfoliated (red) and CVD-grown (black) 1L $MoS_2$ with $SiO_2$.**

## 2. Sensitivity of laser heating to thermal conductivity of supported 2D film

Figure S2 shows the *simulated* temperature rise in 1L $MoS_2$ on $SiO_2$(90 nm)-Si vs. laser spot size, for different values of TBC and in-plane $MoS_2$ thermal conductivity ($k_{2D}$). The figure illustrates the (in)sensitivity of the laser heating experiment (with varying spot size) to the in-plane thermal conductivity. It is evident that for TBC > ~ 5 $MWm^{-2}K^{-1}$ and spot sizes > ~ 300 nm, as in the experiment discussed in the main text, the measurement is insensitive to the thermal conductivity of a monolayer transition metal dichalcogenide (where $k_{2D}$ < ~ 100 $Wm^{-1}K^{-1}$),[2-3] and therefore its $k_{2D}$ cannot be reliably extracted. The $k_{2D}$ values shown in Figure S2 are chosen to be in the range of those previously measured for 1L $MoS_2$ (see e.g. Ref. 2) and are varied in the simulation by more than an order of magnitude to examine the (in)sensitivity of our measurement to $k_{2D}$.

For high thermal conductivity 2D materials such as graphene and hBN, where $k_{2D}$ is ~ 10× larger compared with $MoS_2$,[4-8] the thermal healing length (see Results and Discussion Section in the main text) could be ~ 3× longer (if the TBC is similar). In this case the sensitivity of the measurement to $k_{2D}$ improves, but not significantly. A more reliable extraction of $k_{2D}$ requires that the thermal healing length should be few times larger than the laser spot size, which can be obtained by suspending the 2D film.

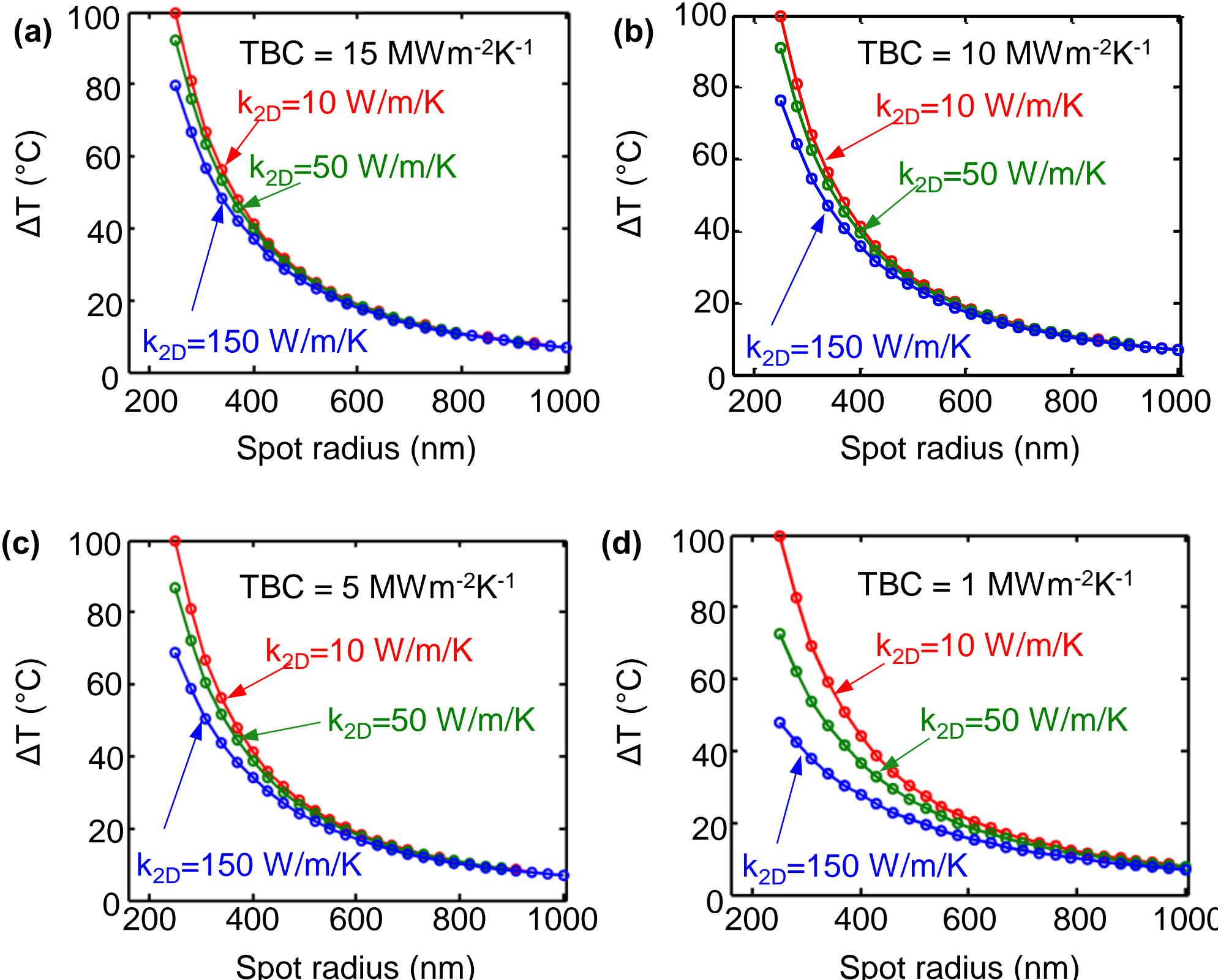

**Figure S2 | Sensitivity of laser heating to thermal conductivity of supported 1L $MoS_2$.** Simulated temperature rise in 1L $MoS_2$ on $SiO_2$(90 nm)-Si *vs*. spot radius with different values of TBC = **(a)** 15 **(b)** 10 **(c)** 5 and **(d)** 1 $MWm^{-2}K^{-1}$ and thermal conductivity $k_{2D}$ = 10 (red), 50 (green), and 150 (blue) $Wm^{-1}K^{-1}$ of the $MoS_2$. The absorbed laser power is normalized to induce $\Delta T_{max}$ = 100 °C in each panel.

## 3. Laser spot

The Raman thermometry technique was historically extended from bulk samples[9] to suspended 2D films,[10] which required varying the spot size in order to extract two fitting parameters: the in-plane thermal conductivity of the 2D film ($k_{2D}$) and the thermal boundary conductance (TBC) at the edges where the film is supported.[11]

Our measurement of fully-supported 2D films, however, is sensitive almost entirely to a single parameter: the TBC (see Figure S2). It is therefore, in principle, possible to extract the TBC from a single measurement. However, the laser spot size dominates the uncertainty of the measurement and it is useful to vary the spot size during the heating experiment, e.g. by sweeping the z-position of the objective.

Figure S3 displays our characterization of the laser spot shape and size. We carry out the knife edge experiment[11-12] illustrated in Figure S3a, where the laser beam is scanned along the x-axis across a sharp edge that blocks the Raman signal, such as a nearby metal film (> ~ 20 nm thick).

The Gaussian shape of the laser intensity yields a decay of the integrated Raman signal area following the form of the complementary error function (*erfc*) as the laser is scanned across the metal edge (Figure S3 b-c). The spot radius can be extracted by fitting an *erfc* to the measured integrated area (Figure 3c). We repeat the knife edge method for varying offsets in the z-axis to find the focal plane and the minimum spot size (shown in Figure S3b-c after correction to set $r_0$ at z=0). We find that in a single measurement the spot size could vary due to small offsets in the z-axis from the focal plane and we carry out the laser heating experiment at varying z-offsets to reduce this uncertainty in our measurement. We note that $r_0$ at the focal plane (z=0) is defined here as in Ref. 11 via $I \propto \exp(-r^2/r_0^2)$, where $I$ is the Raman intensity. Our $r_0 = \sqrt{2}/3\ s_{rad} \approx 0.5 s_{rad}$, where $s_{rad} = 3\sigma$ of a Gaussian profile $I \propto \exp(-r^2/2\sigma^2)$. The measured $s_{rad} \approx 600$ nm and the diffraction limited spot radius is 0.61λ/N.A = 540 nm.

The good agreement between the measured area and the fitted *erfc* shown in Figure S3c confirms the Gaussian shape of the laser at varying offsets in the z-position. We extract the spot radius at each z-position and plot the extracted spot radii *vs*. offset in z to find the minimum spot size as shown in Figure S3d. We note that for each power input of the laser heating experiment we sweep the z-offset to reduce the uncertainty in the laser spot size. The uncertainty in the laser spot size is 10% of the spot radius (shown in Figure S3d), resulting in 20% error in spot area. This uncertainty in spot area is a leading factor in the uncertainty of the measured TBC (see main text). It is also evident that the thermal healing length of 1L $MoS_2$ on $SiO_2$ (where G ~ 15 $MWm^{-2}K^{-1}$ and $k_{2D}$ < ~ 100 $Wm^{-1}K^{-1}$ following refs 2-3, 13) is comparable to the uncertainty in the spot size, further signifying the insensitivity of the measurement to the in-plane thermal conductivity within that range ($k_{2D}$ ~ 100 $Wm^{-1}K^{-1}$), as discussed above.

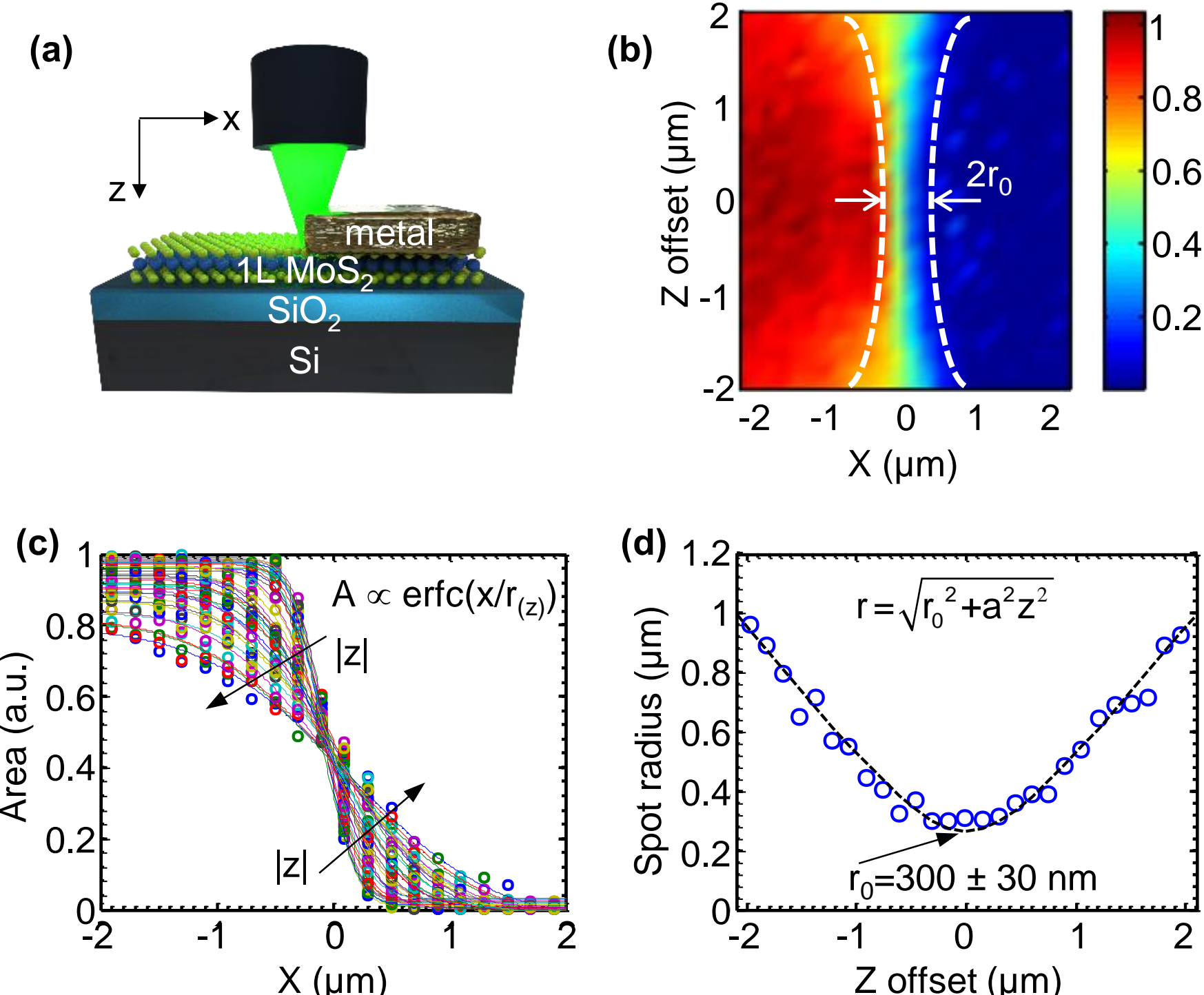

**Figure S3 | Laser beam shape and spot size. (a)** Schematic of the “knife edge” experiment[12] with varying spot size. The laser is scanned in the x-axis across a sharp edge of a metal that blocks the

Raman signal. The spot size is varied by defocusing the laser spot (offset in the z-axis). **(b)** x-z map of the $MoS_2$ integrated Raman signal intensity (normalized). White dashed line is a guide to the eye representing the bounds (horizontal axis) of the spot size for each z-offset. **(c)** The measured integrated area vs. position along the x-axis and fit to complementary error function (*erfc*) for different offsets from the focal plane in z (same data as in b). **(d)** Extracted spot radii (blue symbols) from the *erfc* fits shown in (c) as a function of the offset in z. Dashed black line is a fit to the spot radius dependence on z, with extracted minimum radius $r_0 = 300 \pm 30$ nm. The parameter $a = 0.48 \pm 0.02$ extracted by the fit in (d) is proportional to the N.A (=0.6) of the objective.

## 4. Heating by absorbed power in Si substrate

Thanks to the material selectivity of the Raman technique, we obtained the temperature rise in the Si substrate during the laser heating experiment (see for example Figure 4c in the main text). The Si substrate heats due to absorbing a fraction of the incident laser power, rather than being heated by the $MoS_2$. As a result, the $MoS_2$ film is also further heated by the Si substrate as illustrated in the simulation shown in Figure S4. It is therefore important to consider the measured Si temperature in the analysis of the TBC. The absorbed power in the Si substrate is modeled as a heat source with a Gaussian beam shape in the radial direction and exponentially decaying intensity (Beer-Lambert law for the absorption) in the z-axis with an absorption depth of 0.65 μm.[14]

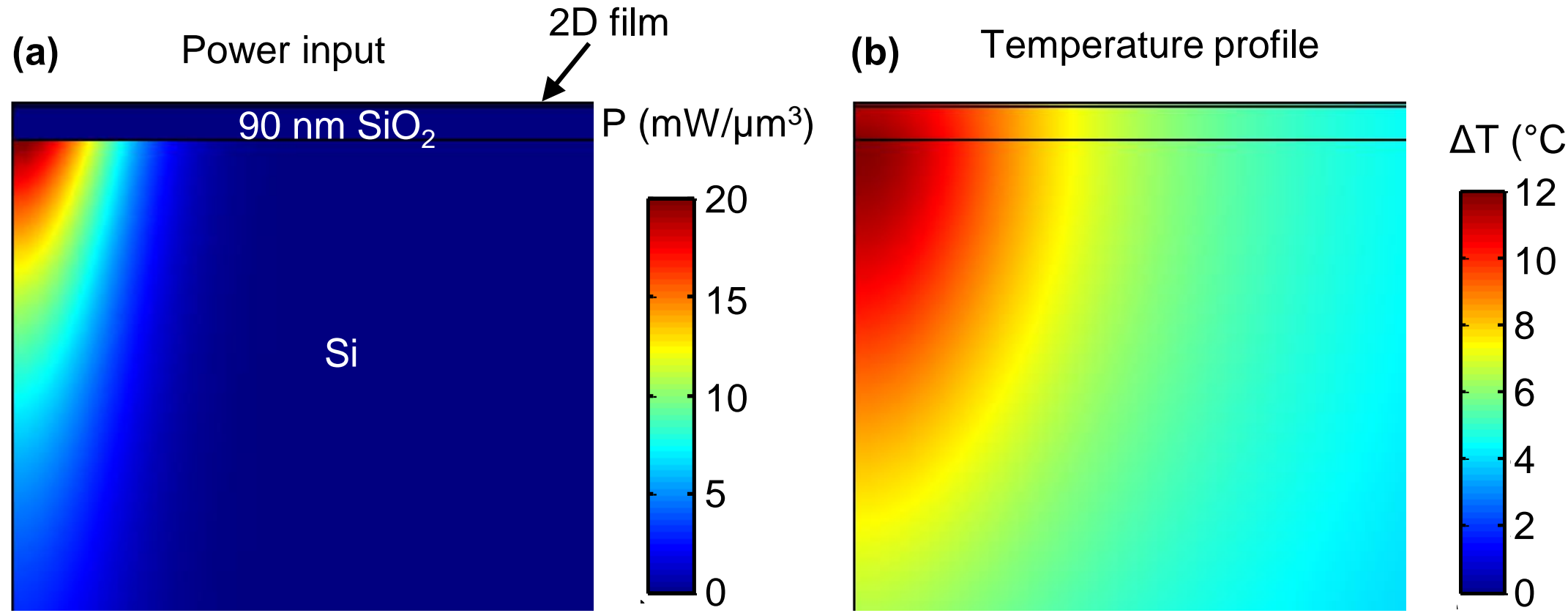

**Figure S4 | Absorbed power and heating of Si substrate. (a)** Simulated absorbed power density in the Si substrate during laser heating experiment. The total absorbed power is 2 mW, the laser spot size at the Si surface is 300 nm and the absorption depth is 0.65 μm.[14] To highlight the effect of heating from the Si substrate, no absorbed power is assumed at the 2D film in this simulation. **(b)** Simulated temperature profile for the power input described in (a), for which only the Si substrate absorbs power. The temperature rise at the 2D film as a result of the substrate heating is similar to the temperature rise at the top surface of the Si with some heat spreading in the $SiO_2$.

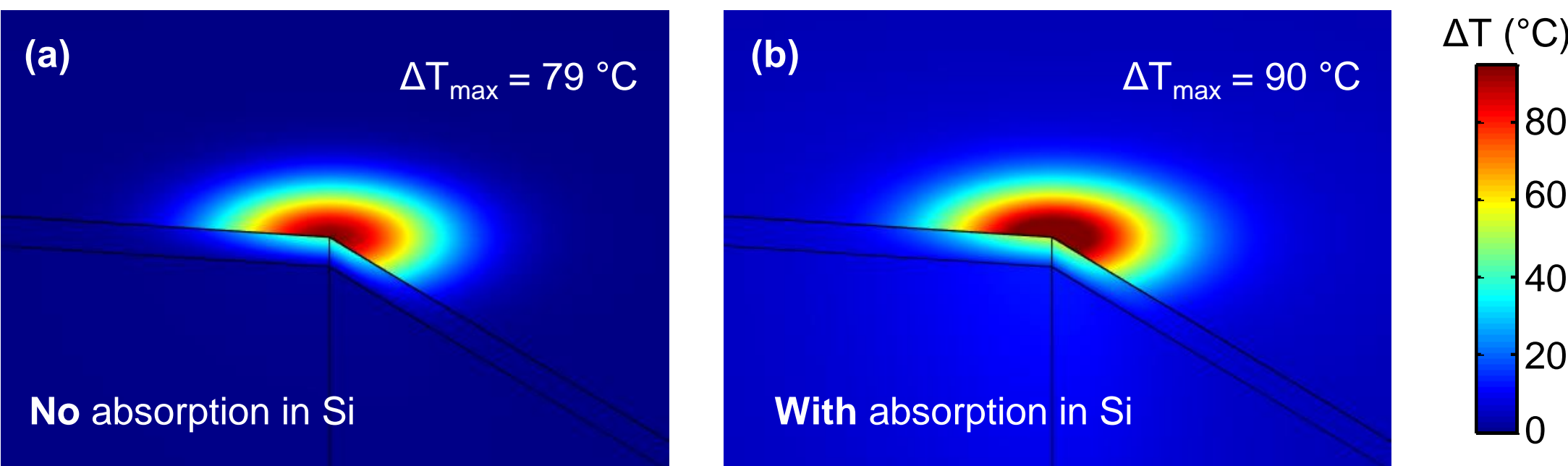

**Figure S5 | Temperature rise of $MoS_2$ with *vs*. without heating from Si substrate.** Simulated temperature rise (ΔT) during laser heating experiment with absorbed power in **(a)** the $MoS_2$ only (no absorbed power in the Si substrate), and **(b)** both $MoS_2$ and Si. It is evident that the absorbed power in the Si substrate heats the $MoS_2$ film (also shown in Figure S4), and therefore the measured temperature of the heated Si substrate must be used in the analysis of the TBC.

## 5. Thermal resistance of incident *vs*. absorbed power

Figure S6 compares the temperature rise in the laser heating experiment of 1L $MoS_2$ with respect to the incident and absorbed laser power. The thermal resistance of the 1L $MoS_2$ (slope) seemingly increases with temperature when considered with respect to the incident laser power. However, when the temperature-dependent absorption is taken into account, the thermal resistance with respect to the absorbed laser power slightly decreases, indicating that the TBC slightly increases with temperature as reported in Figure 4 of the main text.

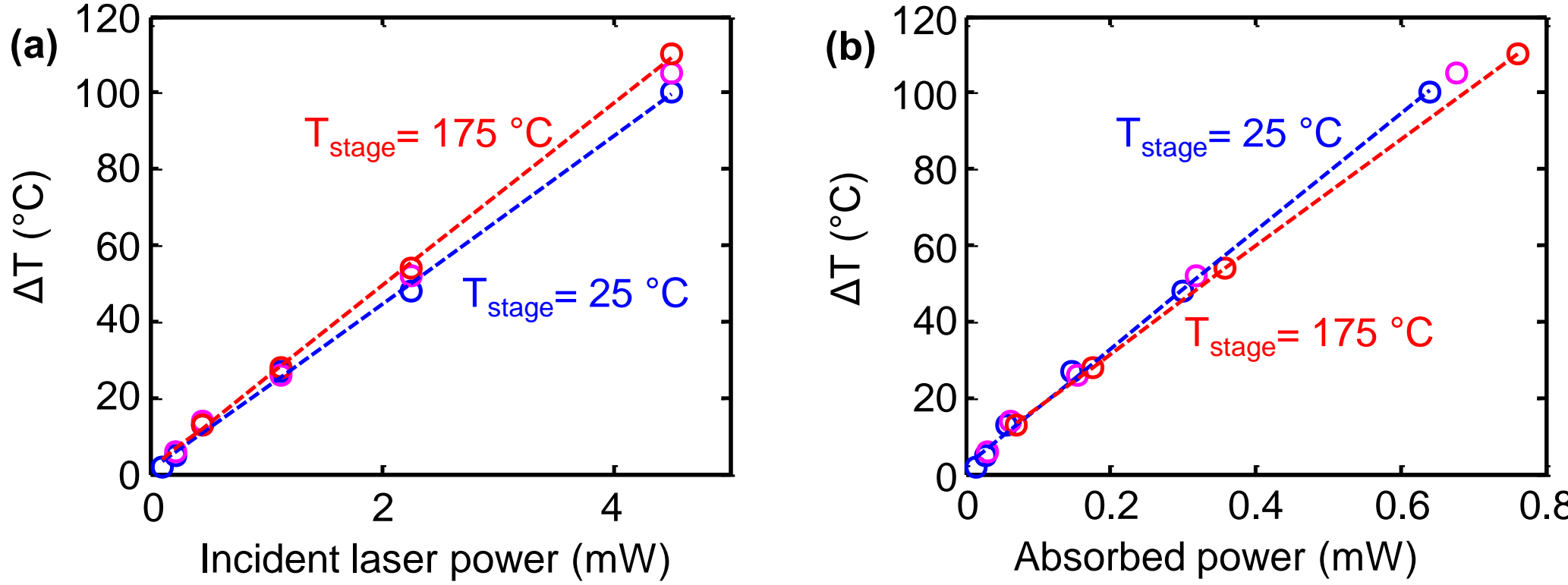

**Figure S6 | Thermal resistance with respect to incident *vs*. absorbed laser power.** Measured temperature rise in CVD 1L $MoS_2$ on $SiO_2$(94 nm)-Si at varying stage temperatures: 25 (blue), 100 (purple), and 175 °C (red) with respect to **(a)** incident, and **(b)** absorbed laser power. The comparison shows that the thermal resistance of the 1L $MoS_2$ (slope, dashed line) *increases* with temperature with respect to the *incident* laser power, but slightly *decreases* with temperature with respect to the *absorbed* power. The different trend is obtained due to the increased optical absorption of the $MoS_2$ at 532 nm with temperature, see Figure 3b in main text.

## 6. Enhancement factor of AlN-Si substrate

The intensity of the electric field of the electromagnetic wave from the laser on a given *substrate* relative to the intensity of the *incident* electric field (of the electromagnetic wave from the laser), termed here the enhancement factor is required to obtain the absorbed power in the supported $MoS_2$ film, and its calculation for $SiO_2$-Si is shown in Figure 3 of the main text. We use transfer matrix method for the calculations. Figure S7 below shows the enhancement factor of the AlN(185 nm)-Si stack *vs*. wavelength. The value 0.91 is obtained at 532 nm (shown in green) and was used for the calculation of the absorbed power in the 1L $MoS_2$ on AlN-Si shown in Figure 4 of the main text.

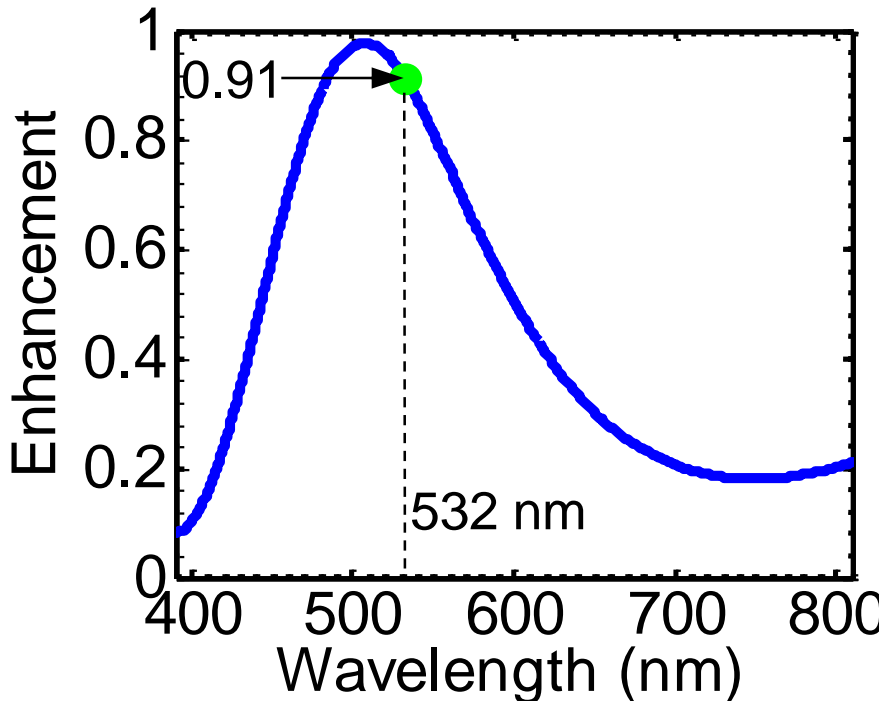

**Figure S7 | Calculated enhancement factor of AlN(185 nm)-Si substrate**. Calculated enhancement factor of AlN(185 nm)-Si substrate vs. wavelength. The value of 0.91 for wavelength 532 nm (laser line used in this study) is indicated and was used to calculate the power absorbed by the 1L $MoS_2$ on AlN(185 nm)-Si substrate shown in Figure 4 of the main text.

**Supporting Information References:**